\documentstyle[12pt,epsfig]{article}
\title{{\normalsize{\bf{High energy neutrino beam generation based on crystal optics.
 }}} \\
\centerline{ \normalsize{Yu. A. Chesnokov and V.A. Maisheev}} 
\centerline{ \normalsize{NRC Kurchatov Institute - IHEP, 142281, Protvino, Russia }} }
\date{}

\begin{document}
\maketitle 

\begin{abstract}
The problem of creation of high energy neutrino  beams on the basis of modern and future circular  proton  
accelerators with the help of traditional technology seems to be expensive and difficult.
Because of this, we propose the solution of this problem based on the usage of
focusing bend single crystals. In the paper we demonstrate the possibilities of acceptance and focusing
of a pion beam with the help of a crystal optical lens system.  As an illustration of these features    
the calculated neutrino fluxes for energy of circulating proton beam equal to 6.5 TeV are presented.   
\end{abstract}

\section{Introduction}
The development of experimental physics of elementary particles is accompanied by the creation of accelerators with ever higher energy.
At present, LHC is the accelerator with the highest proton energy.
The previous high energy accelerators are the SPS (CERN) and Tevatron (Fermilab).
It is well known that such accelerators can be used in two modes.
The first mode is a collider mode when two independent accelerated beams (moving in the opposite 
directions) collide inside the vacuum chamber at special intersection points.
The apparatus standing around this points allow one to obtain information about proton-proton
( or nuclear-nuclear) interactions.
The second mode is the production of secondary beams (such 
as beams of pions, muons , electrons, neutrinos and others).
Experiments using
such beams (fix target experiments) also make it possible to obtain important information about
interactions of various high-energy particles.

Until now, the LHC operates only in collider mode.
However there is the project, in which
the possibilities of the creation of  secondary hadron beams for some experimental program [1, 2] are considered.
In the project the authors assume to use bent single crystals for extraction of proton
beam from LHC.

Besides, in the papers \cite{mch,mch1}   another possibility to obtain secondary beams was considered. It is based on the idea
to produce secondary particles on the target located inside the vacuum chamber of the accelerator.
These particles are extracted from the accelerator with the help of  special focusing crystals.
The focusing crystals can accept secondary particles in a wide angle range, and, hence,  to extract the secondary beam
from accelerator. Additionally, the focusing crystals can focus secondary beam on a target of experiment.  
In Ref. \cite{mch1} for conditions of LHC the example of beam line of positive secondary particles was presented. 
The authors estimate the total length of this beam line about 250 meters.

Neutrino interactions is one of central directions in study of particle physics. 
In particular, these processes are investigated on proton circular accelerators with
the use of special formed neutrino beams
The neutrino beams are the result  
of decays  as $\pi^{\pm} \rightarrow  \mu^{\pm} + \nu (\hat{\nu})$. The special devices are used for increasing flux of a neutrino beam. 
 These devises (magnetic horns) focus parent particles ($\pi$ and $K$-mesons)
into beam close to parallel one. Thus generation of the neutrino beam requires the several stages:
1)  to create the pion beam with the help of the proton beam interacting with a target; b) to accept effectively the pion beam
into a  lens device; c) to transform the pion beam into approximately  a parallel beam;
d)to allow the pion beam to decay in reaction $\pi^+ \rightarrow \mu+\nu$; e) to deflect 
the background particles from a direction of neutrino beam propagation. 
The solution of such  problems in the range of TeV energies of accelerated protons on the basis of 
traditional technology seems to be expensive and complex.  

Because of this, we propose the solution of the problem based on the usage of
focusing bend single crystals. In the paper we demonstrate the possibility 
creation of high energy neutrino beam on the basis crystal optics elements.

This paper is devoted to a study of possibilities of creation neutrino beams on circular proton
supercolliders. For this aim we propose to use special focusing bent single crystals (see \cite{WS} and
 literature therein). The main purpose of our study is the demonstration of reality
of obtaining intensive enough neutrino beams with a wide energy spectrum on the basis of
focusing bent single crystals.

The paper is organized as follows. First, we demonstrate progress in manufacture of the
focusing crystals and discuss the method which allow one to accept valuable part of secondary
particles and to transform  this flux into a practically parallel beam. In the
next section we present calculated energy spectra of pions which can be obtained as result
interaction of 6.5 TeV proton beam with the target. After this we find the neutrino fluxes
emitted within several given angles. Then after short discussion the conclusion follows.

\section{Focusing crystals as a tool for obtaining parallel pion beam}

The first measurements of the beam focusing effect were
performed in the 1990s\cite{MAG,MAG1}. Since then the focusing devices have been  significantly improved\cite{WS}. 
Fig. 1 illustrates the operation principle of such devices.
The 
focusing crystal is  represented by a sum of rectangle
ABCF and triangle FCD (see Fig.1a). Positively charged particles entering the bent crystal in channeling regime
are deflected through the same angle over the distance BC
(AF). For a sufficiently large deflection angle, the channeled
and nonchanneled particles (background) are spatially separated.
The triangular part of the crystal deflects particles
with different transverse coordinate $x$ according to a linear
relationship between the angle and coordinate. Therefore,
the particle trajectories converge at some (focal) point.
The results of recent study of strip focusing crystals one can find in the paper \cite{WS}.

Fig. 1b illustrates the inverse
case of focusing, when the point-like beam from point O
(on a distance $L$ equal to focal length $L_f$ from the crystal
$L \approx L_f$ ) is transformed into practically a parallel beam.
In this case the beam with a small size (in bending plane)
and with a valuable angle divergence may be transformed into
parallel one. The paper \cite{mch1} contains the theoretical description of
the inverse focusing and Ref. \cite{IF} is devoted to the experimental
observation of this focusing mode. 

Taking into account  the importance of inverse focusing for our  study
we  will consider more detail this case  (see Fig. 1c). In Fig. 1c, for clarity, only the triangular part of the
focusing crystal is shown. Here thick (black) circular arcs represent the planar crystallographic channels
and the thin (red) arcs  correspond to centers of these channels. It is easy to see that these
arcs are the projections of crystallographic planes on the plane of bending. The thin (red) lines (defined by $A_1, B_1, C_1$ letters )   are 
perpendicular to the surface of crystal. The straigth lines as
AO, BO, CO are the tangents to corresponding arcs located on a surface of the linear cut.   
There are two angles shown in Fig. 1c. They are $\omega_1$ and $\omega_2$. Obviously, that
$\omega_1=|BB_1|/R $ and $\omega_2= X/L$, where $|BB_1|$ is the the length of arc $BB_1$, $R$ is the bending radius 
of the crystal and $L$ is the distance between crystal and $O$-point. Fig. 1d illustrates the 
crystal in Cartesian system of coordinates before its bending. In this system for the line CD   the transverse coordinate $x$ is
connected  with the longitudinal coordinate $z$  by the equation $z=kx$, where $k$ is a constant coefficient.   
It is obvious that $\omega_1=\omega_2$. From here we found that $L=R/k$ and we see that the  $L$ value is
independent of $x$ and $z$ coordinates. The distance $L$ is practically equal to the focal length $L_f$ of a bent
single crystal (see Ref. \cite{WS}):
\begin{equation}
L=L_f \approx R/k
\end{equation}

We see also that any positively charged particle  emanating from a point $O$ within the maximal and minimal
angles $\omega(x)$ ($\omega_{2, max}=d/L$, $\omega_{2,min}=0$, $d$ is the transverse thickness of crystal )
moves along the straight line and enter in the crystal practically under zero angle relative to crystallographic 
planes. It means that the positive particle may be accepted in a channeling regime. Thus, the accepted  channeled particles
  are formed into a parallel beam. This process was studied in details in the paper \cite{mch1}.
From this study follows that the total efficiency of transformation of the beam into parallel is equal 
to the product $w_T=w_o w_c$, where $w_o$ is the geometrical efficiency  (probability) for a particle emitted from 
point $O$ to be on the surface of the cut in limits of a critical angle of channeling $\theta_c$ relative to the direction of 
crystallographic planes and $w_c$ is the probability (the efficiency) for a particle moving under angle 
less  than $\theta_c$ relative to crystallographic planes to be captured in channeling and to conserve
this statement up to exit from the body of bent crystal. According to the paper \cite{mch1}
the efficiency $w_o$ is maximal when the distance from the point $O$ to crystal is equal to the focal length $L_f$
of this crystal and at this condition it is equal to
\begin{eqnarray}
w_o= 1,\,\, \verb|if| \,\, s_m \le L_f\theta_c, \\
w_o={L_f\theta_c\over s_m}, 
\,\, \verb|if| \,\, s_m \ge L_f\theta_c, 
\end{eqnarray}
where $s_m$ is the half size of particle source (in the point $O$).
We see that at small sizes of particle source the geometrical efficiency $w_o$ is equal to 1.
If $s_m \ge L_f\theta_c$ then only particles from the area with coordinates from $-s_m$ to $s_m$ may be 
captured in channeling regime.

The probability $w_c$ depends on the partition of captured particles in channeling regime and losses of particles     due to  dechanneling in a bent single crystal. For simple one periodic crystallographic planes   
and for large enough bending radii (in comparing with the critical radius $R_c$)  the 
partition of captured particles is equal approximately 0.75 \cite{mch1}. Dechanneling was considered also
in many papers (see, for example \cite{BKC}).

At proton circular accelerators the neutrino beams obtain from decays mainly of $\pi$ and $K$ mesons.
Beams of $\pi$ and $K$ mesons produced on a special target using extracted from
the accelerator proton beam.  Neutrino  flux from $K$ mesons is about 5\% of total flux{\cite{Bon}. In this paper 
we will consider only $\pi^+$-mesons for obtaining neutrino beams.

\section{Schemes of obtaining of $\nu$-beams}
We will consider two different schemes of obtaining neutrino beams with help of focusing crystals at high energy
proton accelerators ( at the LHC accelerator, for example).

At the first scheme we assume that it is  possible to extract a proton beam from an accelerator (for example, with the 
usage of usual strip bending crystals). After this, the  proton beam are focused with the help of conventional magnetic  or crystalline lenses in some point outside an accelerator vacuum chamber.   The target (for pion beam production)  should be placed in this point  (for example, target from beryllium with the length about several tens of centimeters).
The required sizes of the proton beam should be less than $L_f \theta_c $ in the both  transverse planes.

Produced in this target secondary particles move in the direction
of two focusing crystals operating (in mutually orthogonal planes) in inverse mode.
These two focusing crystals placed on the distance equal to their focal lengths and they accept and transform beam of the secondary particles into parallel one.   

At the second scheme we propose to use  the target for production of pions located in the accelerator chamber of 
accelerator.  Fig. 2 illustrates this scheme. The proton beam interacts with target T.  The two focusing
crystals accept the secondary beam and one of them (HFC for definiteness) deflect the secondary beam 
 on some angle. This solution allows one to conserve circulating regime of protons in accelerator and gives the possibility to use the multi-turn passage protons through the target. Really HFC crystal accept particles only
with angles more then $\phi$-angle relative to the direction of motion of the proton beam in an accelerator. 
The full angle acceptances of HRC and VRC are approximately equal to angles $\alpha_h$
and $\alpha_v$, correspondingly. Note that in the vertical plane the accepted  angle 
is symmetrical relative to the direction of proton motion ($\pm \alpha_v/2$) 

After HRC and VRC system the beam of positively charged secondary particles (including pions)  will be approximately parallel (in the both planes). 

However, due to the oscillatory motion of the particles during channeling, a small divergence of the pion beam must take place.
The envelope of secondary beam  with the energy $E$ in horizontal plane is described by equation:
\begin{equation} 
x_{rms}^2(l) = x_{rms}^2(0)+\theta_{rms}^2 l^2
\end{equation}
where $x_{rms}(l)$ is the rms of coordinate of beam distribution on a distance equal to $l$ after crystals and
$\theta_{rms}$ is the rms of angles of beam after crystals. In paper \cite{WS} authors suggest
$\theta_{rms}^2 =\theta_c^2/3$. The analogous equation is valid for vertical plane.

After long enough decay distance the beam secondary particles should be deflected on some angle
  relative to direction of propagation of neutrino beam.  

\section{Spectra of pions}
For calculations of pion spectra in proton-nuclear interactions we used the simple empirical formula \cite{Mal}.
The formula is based  on a fit of precise measurements of particle production by 400 GeV/c protons on 
beryllium targets. In Ref.{\cite{Mal}} value $dN^2_\pi/(dpd\Omega)$ (where $p, \Omega$ are the secondary particle momentum and solid angle)  is calculated for the 500 mm beryllium target. 

In this paper the target production efficiency is also presented:
\begin{equation}
f(L_T)= {\exp(-L_T/\lambda_\pi) - \exp(-L_T/\lambda_p) \over 1-\lambda_p/\lambda_\pi},
\end{equation}
where $L_T$ is the length of the target, $\lambda_\pi$ and $\lambda_p$ are the absorption lengths of
$\pi$ mesons and protons. Note $f(L_T=50 cm) = 0.43$.
Then  we can consider the value $d^2\tilde{N}_\pi/(dpd\Omega)=  d^2N_\pi/(dpd\Omega) /f(L_T)$ as a particle yield per one interacting proton
with a nucleus of target. The value $d^2\tilde{N}_\pi/(dpd\Omega)$ is coupled with the invariant inclusive cross section and 
independent of such parameters as $\lambda_\pi$ and $\lambda_p$ (see Eq. (12) in \cite{Bon}). Thus all the results of calculations of particle yields in our paper must be multiplied
on the efficiency of a target. For a single  passage of beam through the target, the efficiency can be calculated using Eq.(5) and for very thin targets and 
multi passage of a proton beam  through the targets the efficiency can be close to 1.

Fig.3 illustrates the calculated energy spectra of pions for interaction of 6.5-TeV proton beam with a beryllium target. 

The curve 1 is the spectrum of pions  ($d\tilde{N}/dE_\pi$) integrated over the emission angles
   in the horizontal range from 0.1 mrad to 2.1 mrad and in the vertical range from -1 mrad to 1 mrad
relative to the direction of motion of the proton beam.
The curve 2 is the analogous spectrum but in horizontal
range from 0.2 mrad to 2.2 mrad and the same in vertical plane as a previous one. 
The curve 3 is the spectrum when the horizontal and vertical angle range is from -1 mrad to 1 mrad
relative to direction of proton beam motion.

It can be seen that the maximum of the pion spectrum falls on the energy range 150-200 GeV
and  for
energies $\le 100$ GeV  the pion yield is small. As was considered above the pion flux with the help 
of lens objective (see fig. 2) transformed into beam close to parallel one. We can estimate the angle  rms
of this  beam  (neglecting of the pion flux less than 100 GeV). $\theta_{rms} \le \theta_c/\sqrt{3} \approx 12 \mu$rad for the (110) silicon plane and for enough large bending radii.


\section{Calculations of the neutrino fluxes}

The two-particle decay $\pi \rightarrow \mu + \nu$ is considered in many articles (see for example \cite{SH,MP}). 
Here  we will assume in the calculations that the muon neutrino is a massless particle.

Let us denote the neutrino emission angle in a laboratory system with respect to the direction of motion of the pion as $\theta_\nu$.
Then in an ultra relativistic limit and for $\theta_\nu=0$ the neutrino energy becomes
the linear function of pion energy:
\begin{equation}
E_\nu={2Q \over m^2_\pi} E_\pi = 0.427E_\pi ,
\end{equation}
where $Q= (m^2_\pi-m_\mu^2)/2$ and $m_\pi$,  $m_\mu$ are the pion and muon masses.
In the paper we assume that the velocity of light is equal to 1.

Knowing the pion energy $E_\pi$  and the angle  $\theta_\nu$ we can write \cite{MP}:
\begin{equation}
E_{\nu}={Q\over E_{\pi}} {1\over 1-\cos\theta_\nu\sqrt{1-m_{\pi}^2 /{E_\pi}^2}}
\end{equation}
From this equation we can find  two solutions for $E_\pi$:
\begin{eqnarray}
E_{\pi,+}={Q\over E_\nu \sin^2 \theta_\nu} + {\cos\theta_\nu\over \sin^2\theta_\nu} \sqrt{ {Q^2\over E^2_\nu}-m_{\pi}^2 
\sin^2 \theta_\nu }, \\
E_{\pi,-}={Q\over E_\nu \sin^2 \theta_\nu} - {\cos\theta_\nu\over \sin^2\theta_\nu} \sqrt{ {Q^2\over E^2_\nu}-m_{\pi}^2 
\sin^2 \theta_\nu }. 
\end{eqnarray}
In the case of an approximately parallel pion beam  (with the energy distribution denoted as ${dN_\pi \over dE_\pi} (E_\pi)$) 
we  can find the neutrino spectrum (at the condition that we take into account the neutrino only with angles
less than some given  angle $\theta_\nu$) :

\begin{equation}
{dN_\nu\over dE_\nu}(E_\nu)= \int_\Delta {dN_\pi \over dE_\pi}(E_\pi)  F_1(E_\pi) F_2(E_\pi) F_3(E_\pi) dE_\pi,
\end{equation}
where (with $\gamma_\pi=E_\pi/m_\pi$)
\begin{eqnarray}
F_1(E_\pi) =1-\exp{(-l /(L_0\gamma_\pi) )}, \\
F_2(E_\pi)=0.75^2\exp{(-Z_x/l_d(E_\pi))\exp{(-Z_y/l_d(E_\pi))} },\\
F_3(E_\pi) ={\gamma_\pi \beta_\pi\over 2p^*_\nu}. 
\end{eqnarray}
In Eqs.(10)-(13) the function $F_1(E_\pi)$ describes losses of pions  due to their decays on the length equal to $l$ 
($L_0=7.80$ m is the pion decay length in a rest system),   
the function $F_2(E_\pi)$ presents losses of pions coming in the crystal optical system  (the coefficient is equal to $0.75^2$) and propagating
through the both crystals. Here, $Z_x$ and $Z_y$ are the mean longitudinal lengths 
of the HFC and VFC elements (see Fig. 2). 
$l_d$ is the dechanneling length which is a  linear function of a pion energy. Function $F_3$ is the energy distribution of neutrino at
$\pi \rightarrow \mu +\nu$ decay \cite{SH} in a laboratory system ($p^*_\nu=Q/m_\pi$).  
The limits of integration (denoted as $\Delta$) in Eq.(10) are follow (see Fig.4)
The low limit is always equal to  $E_{\pi,0}= E_\nu/0.427$ (see Eq.6) 
When $E_\pi \le m_\pi /\sin \theta_\nu$ result is equal to sum 
$\int_{E_{\pi,0}}^{E_{\pi,-}} + \int_{E_{\pi,+}}^\infty$
and when $E_\pi \ge m_\pi /\sin \theta_\nu$  the result is $\int_{E_{\pi,0}}^\infty$.

Fig 4 explains the choice of limits. There  are two curves (1 and 2) which were presented accordingly to Eqs.(8)-(9).
for $\theta_\nu=$  0.15 and 0.1 mrad, correspondingly. Let us   consider the curve 2, 
for example. Every point in Fig. 4  corresponds to the pair of values: $E_\pi, E_\nu$
For such pairs from the left  the angle $\theta_\nu < $ 0.1 mrad and for
pairs from the right side $\theta_\nu > $ 0.1 mrad. The area under  the curve 3 is empty, or in the other words, there is no real pairs with the given relation (see Eq. (6)) 
Note that between the curves 1 and 2 are pairs with the angle $\theta_\nu$ in the range 0.1-0.15 mrad. 
The black curve passing through the points $P_1$ and $P_2$ corresponds to the maximal value $E_\nu = Q/(m_\pi \sin \theta_\nu) $ for a given value $\theta_\nu$
(and $E_\nu/E_\pi = Q/m^2_\pi$ is valid for this case). 

\section{Results of calculations of neutrino spectra}
In this section we present the results of calculations of neutrino spectra for two schemes of pion beam generation 
(see section 4).

In Figs.5 and 6 the neutrino spectra are shown
  for the cases when the secondary  beams are produced: a)  on the target placed in the vacuum chamber (see Fig. 2)
and b) on the external target, correspondingly.   
For calculations we take the focusing silicon crystal (with channeling in the (110) plane) with the bending radius equal to 50 m. We also assume that the parameter $k=5$. It is means that 
focal  length of the crystal is equal to 10 m (see Eq.(1)). 
In calculations we set the decay distance for the pion beam equal to 5 km.
Besides, we assume that the transverse sizes of the pion beam on the target in the both planes (horizontal and vertical) are in the accordance
with Eq.(2). Simple estimations show that main neutrino flux arises from pions with energies less than 2000 GeV (see Fig. 3). For the (110) silicon plane 
the critical radius \cite{TS} is equal to $R_c[m] = E_\pi [GeV]/600$. It is equal to  3.33 m
(for $E_\pi=2000$ GeV).  It means that we can take the critical angle of channeling as 
$\theta_{c} [mrad]= 0.207/\sqrt{E_\pi [GeV]}$\cite{BGM}. From here $\theta_c=  4.6\mu$rad (for $E_\pi=2000$ GeV). It means that for effective transformation
of the pion beam its transverse half sizes should be less than 0.046 mm. The transverse size of pion beam determined by size of the proton beam on a target or a size of a target (if this size less than size of proton beam). For comparison we can point out that in
the 1st inter crossing point of the LHC the rms size of the circulating proton beam is
equal to 0.0167 mm. 
For calculations of particle losses due to dechanneling process we use (see Eq.(12)) the linear dependence of dechanneling
length $L_d [cm] = 0.057E_\pi [GeV]$ \cite{BKC}.

Fig. 2 illustrates the second scheme of forming of the parallel pion beam. Here the circulating proton beam
interacts with a thin target (inside of a vacuum chamber). Pions with small angles (in the bending plane)
 relative to the direction of primary proton beam are accepted into the focusing bent single crystal and 
are deflected out from an accelerator chamber. Our calculations were done for the silicon single crystals
(with the (110) planar orientation). We take for calculations $R=50$ m, $L_f=10$ m, FA=5 cm and 0 (for HFC and VFC, correspondingly) and DF=1 cm (see Fig. 1d).     
Thus, the crystal focusing deflector  (HFC) allow one to provide the deflection angle more than 1 mrad.
The efficiency of a thin 
target at multi turn passage of the proton circulating beam throught it may be close to 1.  Because of this, the spectra in Fig. 5 do not required
corrections. 

Concerning of the case of the external target we can point out the follows thoughts.
As was considered in section 4 for obtaining real fluxes of neutrino we should take 
into account  the efficiency of the external target (see Eq. (5)), or 
in the other words, the results of calculations in Fig. 6 should be  multiplied on a coefficient  
about 0.3-0.4.
In the case under consideration we   assume that the proton beam is extracted from accelerator in one 
way or another. It allows us to take short crystals:FA=0, DF=1 cm (see Fig.1 d).

It should be noted in Eq. (10) and for function $F_2$ we use the mean lengths of crystals along $z$-axis.
We believe that this approximation is valid for our calculations but  we present more correct form of Eq.(10):
\begin{equation}
{dN_\nu\over dE_\nu}(E_\nu)= \int_0^{Z_h}\int_0^{Z_v}   {    {dN_\nu\over dE_\nu}(E_\nu, z_h, z_v) \rho(z_h) \rho(z_v) dz_h dz_v },                                                           
\end{equation}
where $\rho(z_h), \rho(z_v)  $ are the normalized per unit  distributions  of pions at the entrance of $HFC$ and $VFC$.
We think that $\rho(z_h) =\rho(z_v)   =1/d $ is good enough approximation.

\section{Discussion}
First of all we would like to note that this is the first study devoted to the use the bending single crystals for focusing pion (kaon) beams and generation of  high energy neutrino beams. Here we demonstrate main principles and tools required to solve this problem. However, we could not consider the various issues in detail.
The Figs 5 and 6 illustrate the calculated energy neutrino spectra, which are main results of our study. 
For comparison in Fig.5 the calculated neutrino spectrum\cite{Bon} is shown for the CHARM II detector\cite{CH2} exposed 
at the CERN-WANF beam \cite{WANF}. This experiment was carried out at the proton energy equal to 450 GeV. 
The measured in this experiment neutrino spectra are in good agreement with the calculations.
Note that the secondary beam was focused with the help of magnetic horns with the angle acceptance
equal to 8 mrad. In Ref. \cite{Bon} calculations presented for $10^{13}$ protons on target (p.o.t.).
We recalculated this on per one proton on target (by dividing on $10^{13}$).

It is useful to find the transverse sizes of the pion beam at the end of the decay length.
In Eq.(4) we can neglect by the first term and then we get  that the rms of pion beam (at the end of 
decay length equal to 5 km) is about 6 cm and it increases approximately linearly as a function of distance.
For this estimation we neglected by the small flux of pions at energy less than 100 GeV.

Note that the angle acceptance (equal to 2 mrad in the both planes) of pion beam cover the transverse
area equal to 2  $cm^2$ on the distance in 10 m from the target (see Fig. 2 ). The experience of fabrication of 
focusing crystals \cite{WS} shows  that the transverse thickness of such crystals should be less than 2-4 mm.
However, the area in about several square centimeters may be covered with the help of several focusing crystals.
We do not see problems in this question.   

Eq. (10) represent approximate analytical  solution  of problem of calculation of neutrino spectra with emission angle less
than some given angle $\theta_\nu$.
This solution is based on possibility to obtain the pion beam practically parallel in the horizontal and vertical
transverse planes. This approximation is valid at condition that the angle rms of the pion beam is significantly
less than $\theta_\nu$. In our study in section 4 we found that   $\theta_{rms} \le \theta_c/\sqrt{3} \approx 12 \mu$rad.
 This result was calculated for $E_\pi=100$ GeV. However, the main contribution in the neutrino flux bring
in several times more energy pions and hence the effective value of angle rms should be less.     

Let us assume that the beginning of the decay tunnel is at the distance $L_1$ and its end at the distance $L_2$ 
from the detecting neutrino device (with the transverse sizes equal to $\pm r_D$ in the horizontal and vertical planes).
Then all neutrinos with emission angles no more than $r_D / L_1$ will enter into the detector, and neutrinos with angles greater than $r_D / L_2$
can not get into it. This simple consideration shows that the knowing of neutrino spectra at different emission angles allow us to calculate
spectra of the neutrino passing through the real detector. 

Now we touch the problem of focusing of negatively charged particles. 
Obviously the solution of this problem allow one to obtain the beam of antineutrino.
The measurements\cite{gu,sc1} performed 
in 120 GeV/c  electron and 150 GeV/c  negative pion beams give for   the dechanneling lengths of (110) silicon plane
 a value equal to 0.6 and 0.91 mm, respectively.  One can expect linear increasing with energy of this value.
So, the results of measurements of dechanneling length  in electron beams with the energy about 10 GeV
give value about 0.05 mm \cite{Wis}. In the experiment \cite{Sc} the maximal  length of focusing crystal was equal to 3 mm along
beam direction. It means that the focusing effect for negatively charged particles may be observed for 
energies beginning from several hundreds GeV. In whole the problem of antineutrino beams requires
careful study.

\section{Conclusion}
In the paper the probability of generation neutrino beams at high energy colliders was studied.
 The focusing bent crystals are proposed to be used as elements forming a parallel pion beam.
For energy circulating protons equal to 6.5 TeV  the neutrino spectra were calculated. 
In general, our research points to the prospects of application of lens systems from single crystals
to generate  high-energy neutrino beams.

\section{Acknowledgments}
We acknowledge
financial support from the Russian Science
Foundation (Grant No. 17-12-01532).


\newpage
\begin{figure} 

\begin{center}
\scalebox{0.8}
{\includegraphics{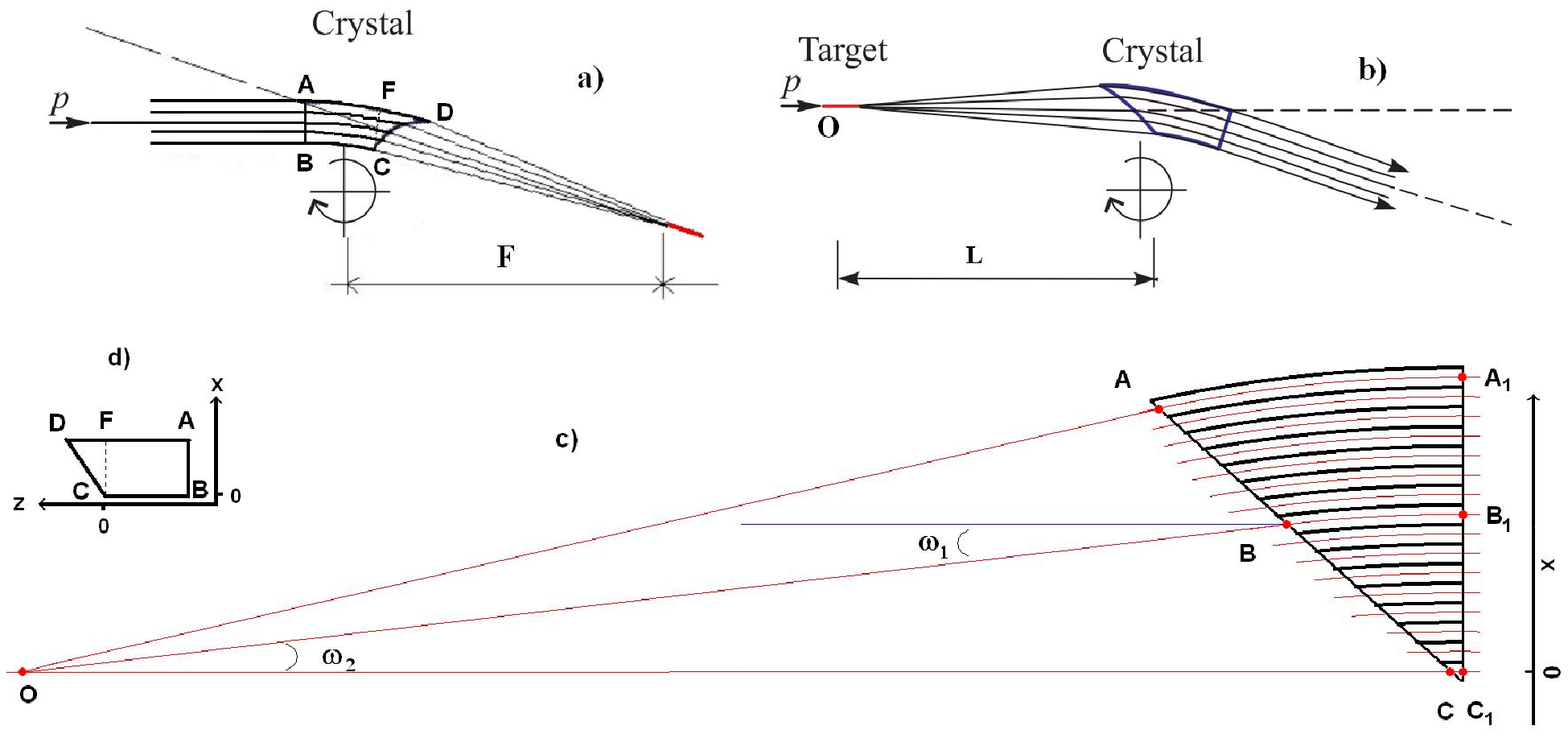}} 

{\caption{
  Focusing bent crystals: a) focusing of parallel beam into point,
b) focusing of point-like beam into parallel, c) the principle of the operation 
of a focusung crystal (for the case b)), d) focusing crystal before installation in
the holder.
}}

\end{center}

\end{figure}

\begin{figure} 

\begin{center}
\scalebox{0.8}
{\includegraphics{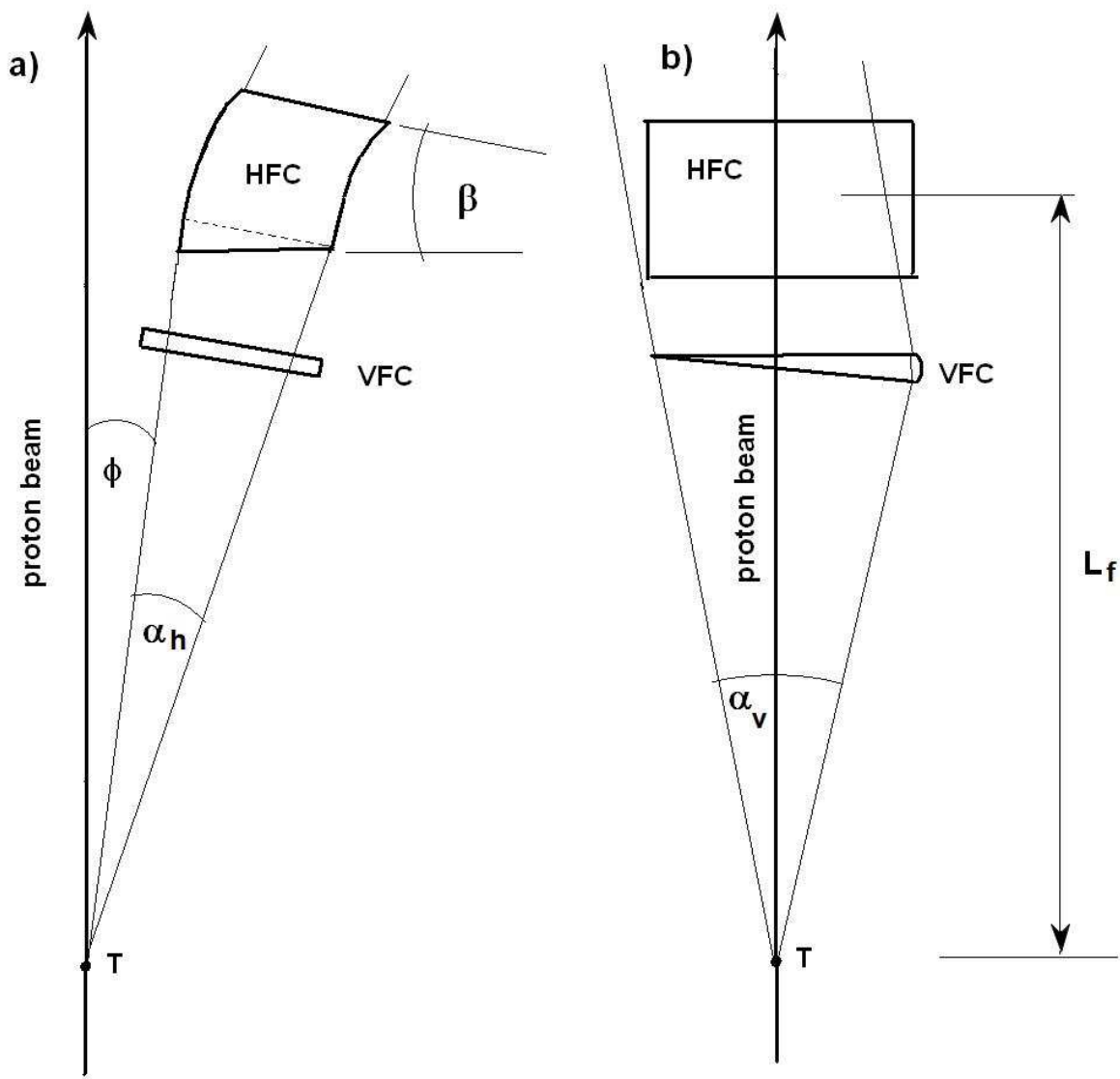}} 

{\caption{
Scheme of generation and formation of a parallel beam of pions:
HFC and VFC are horizontal and vertical bent focusing single crystals, correspondingly,
$\alpha_h$ and $\alpha_v$ are horizontal and vertical angles of acceptance of pions,
respectively, $\beta$ is the bending angle, $\phi$ is angle between the proton beam and   
the edge of the crystal, $L_f$ is the focal length of the crystals $T$ is the target.a) top and b) side view.
}}

\end{center}

\end{figure}

\begin{figure} 

\begin{center}
\scalebox{0.8}
{\includegraphics{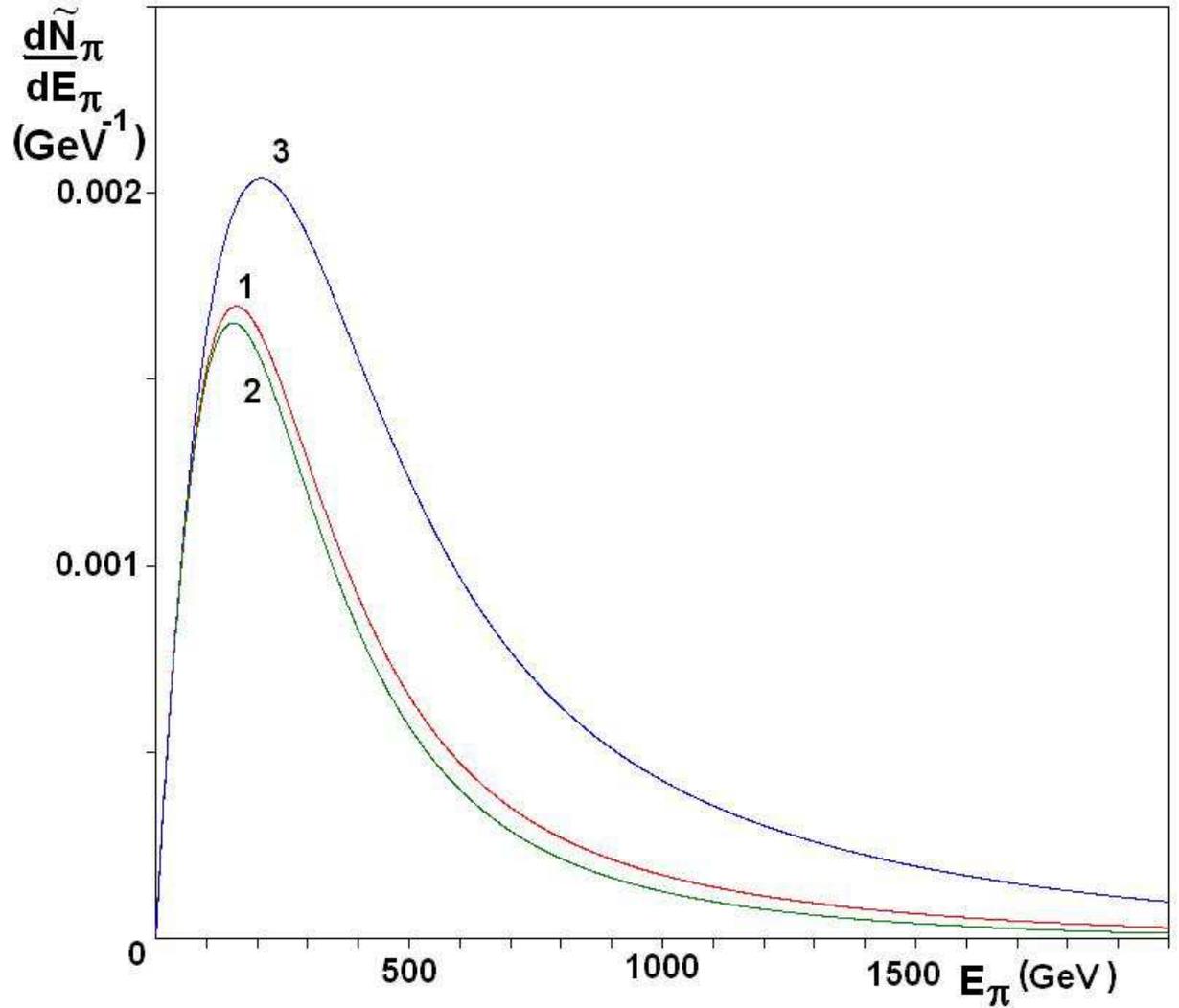}} 

{\caption{
Calculated pion spectra at the entrance in focusing crystals.
The curve 1 and 2 were calculated for $\alpha_h=\alpha_v=2$ mrad and for
$\phi$=0.1 and 0.2 mrad, correspondingly. The curve 3 was calculated at same
$\alpha_h$ and $\alpha_v$ angles but for central passage of pion beam relative
to proton beam.  
}}

\end{center}

\end{figure}

\begin{figure} 

\begin{center}
\scalebox{0.8}
{\includegraphics{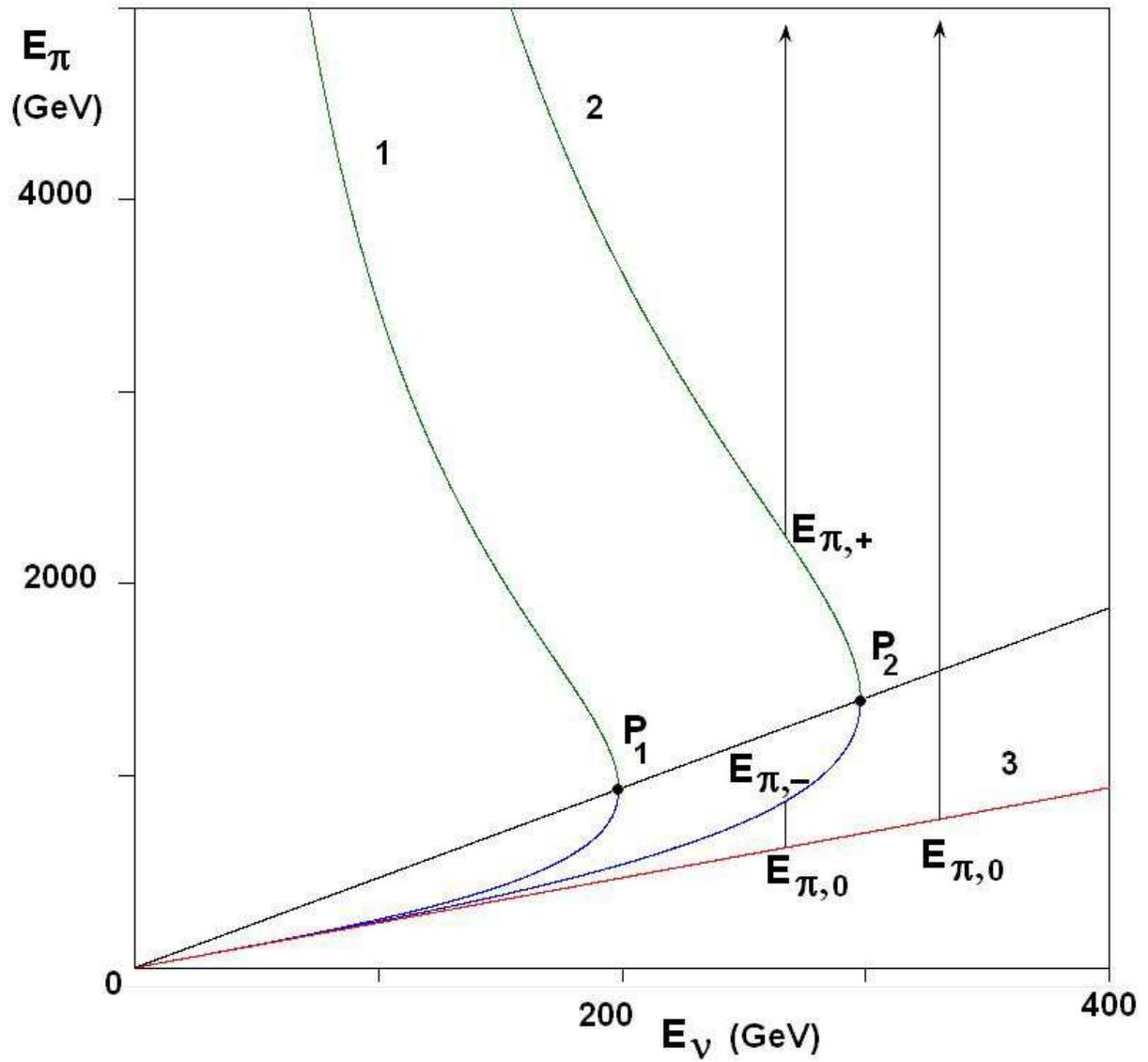}} 

{\caption{
The $E_\nu, E_\pi$ plot. The curves 1 and 2 correspond to  sets of points  
 $E_\nu, E_\pi$ with the emission angles equal to 0.15 and 0.1 mrad, the curve 3
present the couple in the accordance with Eq.(6).
}}

\end{center}

\end{figure}

\begin{figure} 

\begin{center}
\scalebox{0.8}
{\includegraphics{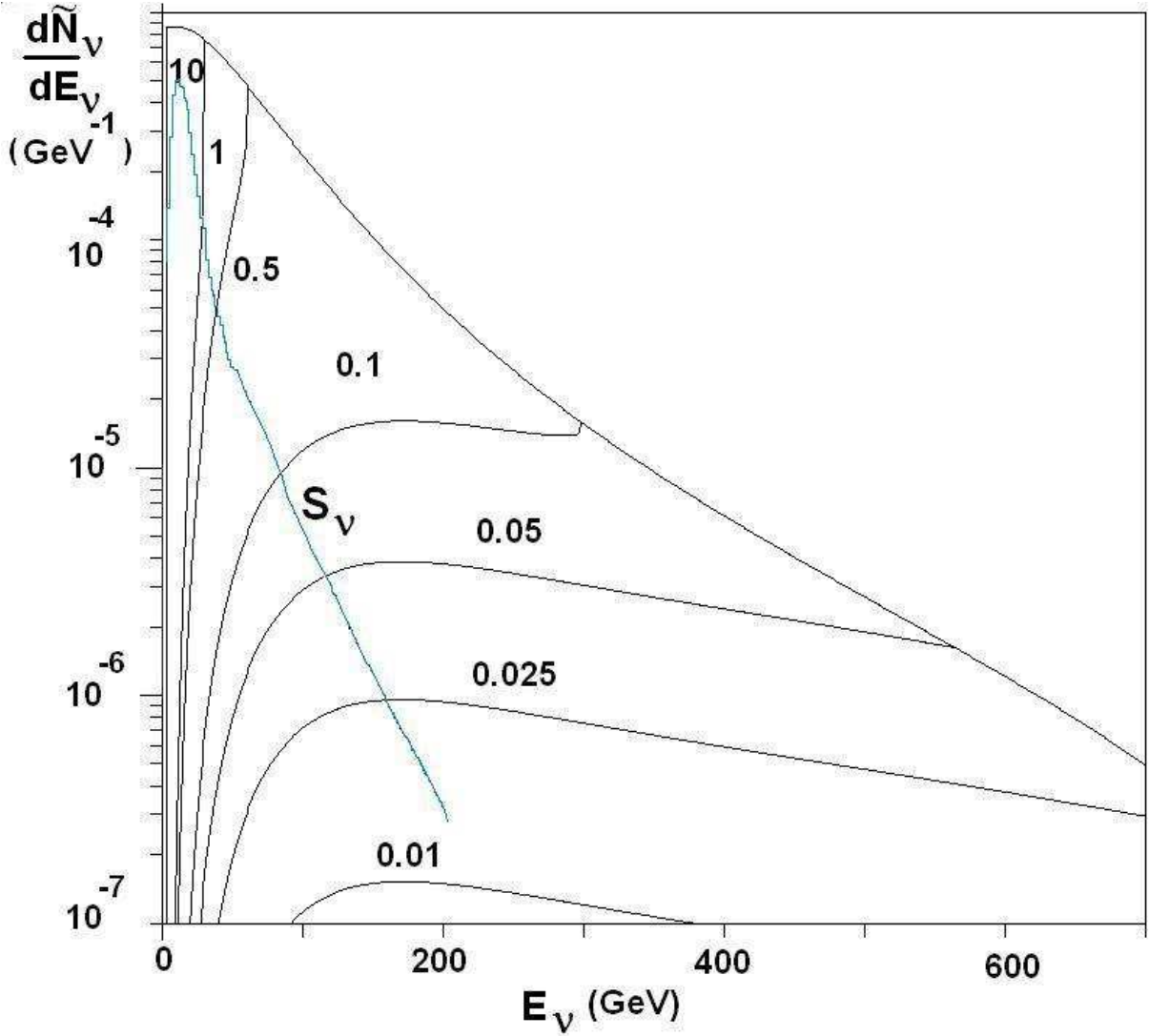}} 

{\caption{
The calculated energy distributions of neutrino beams.
They are calculated for neutrino from zero neutrino emission angle $\theta_\nu$ up to some
its value (the numbers near curves). The proton energy 6.5 TeV. It is assumed that
target is placed in vacuum chamber.
The blue curve $S_\nu$ is the calculations\cite{Bon}.
}}

\end{center}

\end{figure}

\begin{figure} 

\begin{center}
\scalebox{0.8}
{\includegraphics{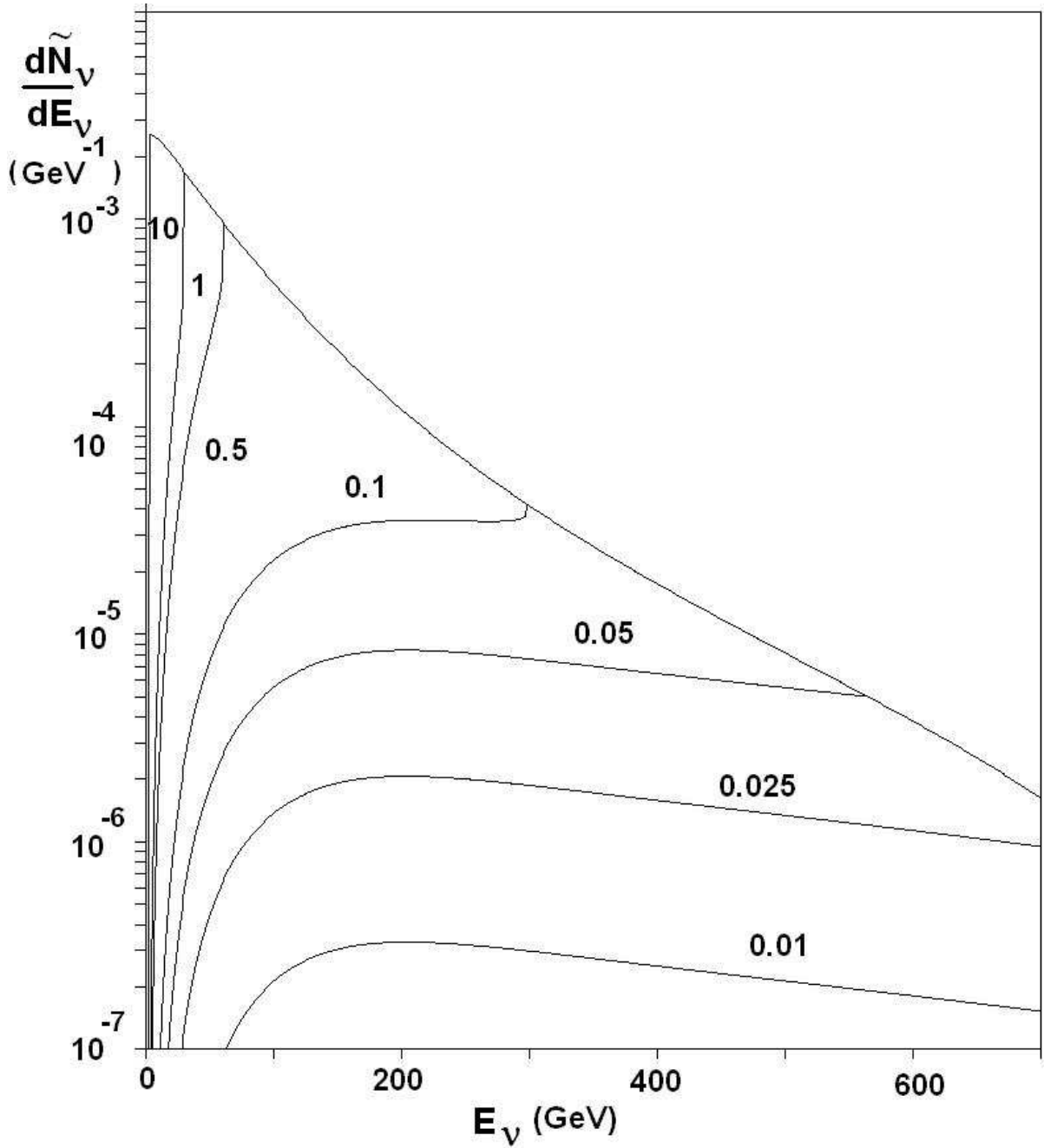}} 

{\caption{
The calculated energy distributions of neutrino beams.
They are calculated for neutrino from zero neutrino emission angle $\theta_\nu$ up to some
its value (the numbers near curves). The proton energy 6.5 TeV.
It is assumed that  target is placed outside  vacuum chamber.
}}

\end{center}

\end{figure}

\end{document}